\documentclass[aps,pra,superscriptaddress,twocolumn]{revtex4-1}
\usepackage{dcolumn}
\usepackage[utf8]{inputenc}
\usepackage{amsmath}
\usepackage{amsfonts}
\usepackage{amssymb}
\usepackage{graphicx}
\usepackage{hyperref}
\usepackage{xcolor}

\newcommand{\ket}[1]{\ensuremath{\left|{#1}\right\rangle}}
\newcommand{\bra}[1]{\ensuremath{\left\langle{#1}\right |}}

\begin{document}

\title{Boosting entanglement generation in down-conversion with incoherent illumination}
\author{Lucas Hutter}
\affiliation{Instituto de F\'{\i}sica, Universidade Federal do Rio de Janeiro, Caixa Postal 68528, Rio de Janeiro, RJ 21941-972, Brazil}
\affiliation{Instituto de F\'{\i}sica, Universidade Federal Fluminense, Brazil}
\author{G. Lima}
\affiliation{Departamento de F\'{\i}sica, Universidad de Concepci\'on, 160-C Concepci\'on, Chile}
\affiliation{Millennium Institute for Research in Optics, Universidad de Concepci\'on, 160-C Concepci\'on, Chile}
\author{S. P. Walborn}
\affiliation{Instituto de F\'{\i}sica, Universidade Federal do Rio de Janeiro, Caixa Postal 68528, Rio de Janeiro, RJ 21941-972, Brazil}
\affiliation{Departamento de F\'{\i}sica, Universidad de Concepci\'on, 160-C Concepci\'on, Chile}
\affiliation{Millennium Institute for Research in Optics, Universidad de Concepci\'on, 160-C Concepci\'on, Chile}

\begin{abstract}
Entangled photons produced by spontaneous parametric down-conversion have been of paramount importance for our current understanding of quantum mechanics and advances in quantum information. In this process, the quantum correlations of the down-converted photons are governed by the optical properties of the pump beam illuminating the non-linear crystal.  Extensively, the pump beam has been modeled by either coherent beams or by the well-know Gaussian-Schell model, which leads to the natural conclusion that a high degree of optical coherence is required for the generation of highly entangled states.  Here, we show that when a novel class of partially coherent Gaussian pump beams is considered, a distinct type of quantum state can be generated for which the amount of entanglement increases inversely with the degree of coherence of the pump beam.  This leads to highly incoherent yet highly entangled multi-photon states, which should have interesting consequences for photonic quantum information science.
\end{abstract}

\maketitle

{\em Introduction.---}Quantum correlated photons generated in the process of spontaneous parametric down-conversion (SPDC) have played a key role in the development of quantum information science over the last decades.  Originating from the conservation of momentum, there is robust entanglement generation between the transverse spatial variables (postion/momentum) of the down-converted photons \cite{howell04}, which has attracted considerable attention as it can be used to define high-dimensional quantum systems \cite{neves05,fedorov07,pires10,dada11,straupe11,edgar12,krenn14,schneeloch19}, as well as engineered in a number of ways \cite{pittman95,monken98b,nogueira04,abouraddy07,gomes09a}. This is due to the fact that the  down-converted photons can inherit properties of the pump laser beam \cite{pittman95,monken98a}, which provides interesting relations between the classical optical properties of the pump field and the non-classical characteristics of the multi-photon state. Thus, SPDC provides rich and flexible quantum state engineering that is crucial for many fundamental studies and applied research.

 Several authors have already studied the spatial entanglement of SPDC in a more generalized framework than usual, where partially coherent pump beam illumination is considered  \cite{lima08,jha10,Giese18,Defienne19,Zhang19,joshi20}.  In particular, Refs. \cite{Giese18,Defienne19,Zhang19} addressed the position and wave vector quantum correlations of down-converted photons produced by a pump beam with partial transverse spatial coherence, described by the well-known Gaussian-Schell model (GSM)\cite{mandel95}. They showed that, in this case, highly entangled states can only be observed when the pump beam has a large degree of spatial coherence.

However, the GSM beam is not the only example of a partially coherent beam. In 1993 Simon and Mukunda introduced the ``twisted Gaussian Schell Model" (TGSM), which is a more general partially coherent Gaussian beam with rotational symmetry around the propagation axis \cite{simon93,simon98}.  This model predicted the existence of novel correlations between the transverse variables of Gaussian beams, a property they dubbed the ``twist phase".  However, the twist phase is not a phase in the usual sense, and in fact vanishes in the coherent limit.  These beams have been experimentally realized \cite{friberg94,ambrosini94}, and recent studies have shown their improved resilience against turbulence-induced degradation effects when compared with traditional GSM beams \cite{wang10,zhou20} with applications in imaging \cite{liu19}. They also carry orbital angular momentum, and therefore can find applications in biophysics and metrology \cite{grier03,dambrosio13b}. 

Motivated by these results, we present here several advantages of adopting TGSM beams for the process of SPDC. In particular, we show that highly mixed, highly entangled states can be produced by exploiting the twist phase in TGSM beams. Counterintuitively, in this case the entanglement actually increases with the incoherence of the pump beam. This effect is a consequence of the infinite-dimension of the quantum states defined in terms of the position/momentum of down-converted photons, a fact that allows for separable and highly entangled states to be arbitrarily close together in state space \cite{clifton99, eisert02,adesso05}.  In addition, we are able to connect the SPDC entanglement with the twist phase of the TGSM pump beam, a novel optical trait in its own right, with useful applications in optics.  This should open the way for producing highly entangled photons in a highly mixed state, which have many potential applications in quantum information science.

{\em TGSM beams.---}The spatial degree of freedom of a paraxial and monochromatic optical field with wave number $k$ can be described by the near-field (position) and far-field variables (wave vector). For a paraxial field propagating in the $z$ direction, let us define the position in a transverse plane as $\boldsymbol{r} = (r_x,r_y)= (x,y)$, and the transverse wave vector as $\boldsymbol{q}=(q_x,q_y)$.   Defining a vector $\xi=(x,q_x,y,q_y)$, the second moments of these variables can be written in a $4 \times 4$ covariance matrix (CM) $V$.
Optical fields with a Gaussian transverse profile are completely and uniquely described by the CM $V$, up to a translation.  The same can be said for Gaussian states in quantum mechanics \cite{adesso07}.
\par
\begin{figure}
\includegraphics[width=.4\textwidth]{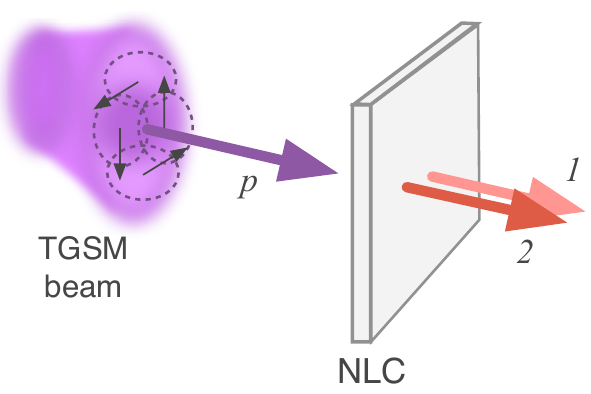}
\caption{A partially coherent TGSM beam is incident on a non-linear crystal producing photon pairs.}
\label{fig:example}
\end{figure}
\label{sec:TGSM}
A TGSM beam is therefore uniquely characterized by its CM  \cite{simon93,simon98}:
\begin{equation}
T=\begin{pmatrix}
\sigma^2 & -\frac{k\sigma^2}{R} & 0 & k u\sigma^2 \\
-\frac{k\sigma^2}{R} &\tau^2 & -k u\sigma^2 & 0 \\
0 & -k u\sigma^2 & \sigma^2 & -\frac{k\sigma^2}{R} \\
k u \sigma^2 & 0 & -\frac{k\sigma^2}{R} & \tau^2 \\
\end{pmatrix}. \label{VMSGTdim}
\end{equation}
Here $\sigma$ is the beam waist, $\tau^2=\frac{1}{\delta^2}+\frac{1}{4\sigma^2}+ k^2 \left( \frac{\sigma^2}{R^2} + u^2\sigma^2 \right)$ is the variance of the wavevector distribution, $\delta$ is the transverse coherence length, $R$ is the radius of curvature. The parameter $u$ is the so-called twist phase, here with dimension given by length$^{-1}$.  These are Gaussian beams in which the position coordinates $x$ and $y$ are coupled to the wave vector coordinates $q_y$ and $q_x$, respectively.  This results in a non-zero angular momentum $\langle L_z \rangle =  2 \hbar k u\sigma^2$, giving origin to the nomenclature "twist".     As particular cases of a TSGM beam, the well-known (rotationally symmetric) GSM \cite{sudol84,mandel95} is obtained by setting $u=0$, and a spatially coherent Gaussian beam is recovered by setting $\delta=\infty$.   The positivity constraints on the CM 
\eqref{VMSGTdim} lead to \cite{simon93,simon98}: $\lvert u \rvert \leq 1/{k \delta^2}$.
Thus, the twist phase tends to zero for a perfectly coherent beam $(\delta \rightarrow \infty)$.  
We note that the twist phase appears in the variance of the momentum coordinates $\tau^2$ through the term $k^2u^2\sigma^2$, which causes increased beam divergence as a function of $u$ \cite{simon93, friberg94, simon98}.  TGSM beams have been produced and studied experimentally, and they can be constructed as a convex (incoherent) combination of coherent Gaussian beams with different transverse phases  \cite{friberg94,ambrosini94}, as illustrated pictorally in Fig. \ref{fig:example}.

{\em Partially coherent SPDC.---} Consider now that a TGSM is used to pump a non-linear down-conversion crystal, as shown in Fig. \ref{fig:example}. Let us assume that the down-converted photons (1,2) are degenerate, so that $k_1=k_2=k/2$, where $k$ is the wavenumber of the pump beam. Let us define the global coordinates:
\begin{equation}
\boldsymbol{q}_{\pm} = \boldsymbol{q}_1\pm\boldsymbol{q}_2;  \hspace{4mm} {\boldsymbol{r}}_{\pm} = \frac{1}{2}(\boldsymbol{r}_1 \pm \boldsymbol{r}_2). 
\label{eq:r+-}
\end{equation}
Under appropriate conditions (see Appendix \ref{supmat}), it is well-known that the two-photon state is nearly separable in these global sum and difference coordinates \cite{monken98a,law04,walborn10,Schneeloch16,Giese18}.  For example, in appendix \ref{supmat} we show that for a partially coherent pump beam, the two-photon Wigner function is  $\mathcal{W}(\boldsymbol{\xi}_+,\boldsymbol{\xi}_-) = W_{+}(\boldsymbol{\xi}_+) W_{-}(\boldsymbol{\xi}_-)$, where  $\boldsymbol{\xi}_\pm=(x_\pm,{q_\pm}_x,y_\pm,{q_\pm}_y)$ are the global phase space coordinates.  Here $W_+$ is the Wigner function describing the spatial properties of the pump laser, and $W_-$ is the Wigner function of the so-called phase matching function $\mathcal{S}$ \cite{walborn10,Schneeloch16}.  Though $\mathcal{S}$ is not a Gaussian function, it can be approximated using a double-Gaussian representation \cite{Schneeloch16}, which in most situations is enough to describe the salient features of the two-photon state \cite{law04,fedorov07,Schneeloch16}.     
\par
The separable form of the two-photon state with respect to the global coordinates allows for several immediate conclusions.  First, the purity the two-photon state is given by $\mu_{12}=\mu_+\mu_-$, where $\mu_{\pm}$ is the purity associated to $W_\pm$ (see Appendix \ref{supmat}).  
The second implication is that the $8 \times 8$ two-photon CM is
\begin{equation}
G = \begin{pmatrix}
V_+ & 0 \\
0 & V_-
\end{pmatrix}, \label{MC+-}
\end{equation}
where $V_\pm$ is the the $4 \times 4$ CM describing second moments of $\boldsymbol{\xi}_\pm$.  
Thus, to study properties of the two-photon state when the pump is a TGSM beam, we simply set $V_+$ equal to  the CM of \eqref{VMSGTdim}.  In this way, the purity $\mu_{+} =[{4 \sqrt{\mathrm{Det}(V+)}}]^{-1} = \beta^2$, 
where $\beta^2=(1+4\sigma^2/\delta^2)^{-1}$ is the dimensionless normalized coherence parameter of the pump beam \cite{friberg94} ($0 \leq \beta^2 \leq 1$).    For a coherent beam we have $\beta^2=1$ ($\delta^2 \gg \sigma^2$), while for a completely incoherent beam $\beta^2=0$ ($\delta^2 \ll \sigma^2$).  Thus, the purity of the two-photon state $\mu_{12}=\mu_- \beta^2$ is proportional to the coherence of the TGSM pump beam.
\par
For the double Gaussian approximation to the phase matching function, we use the diagonal CM $V_-=\mathrm{diag}(\sigma_{-}^2,\Delta_{-}^2,\sigma_{-}^2,\Delta_{-}^2)$, since it is approximately separable in the $x$ and $y$ spatial directions \cite{abouraddy07}. Following Schneeloch and Howell \cite{Schneeloch16},  the variances are  $\sigma^2_- = 9 L/10 k$ and $\Delta^2_- = 3 k/2 L$, where $L$ is the length of the crystal in the longitudinal direction.  
 \par 
To write the two-photon CM $G$ in coordinates describing the individual photons, we define $\boldsymbol{\xi}_{12}=(\boldsymbol{\xi}_{1},\boldsymbol{\xi}_{2})$, with $ \boldsymbol{\xi}_{j}= (x_j,{q_j}_x,y_j,{q_j}_y)$ $(j=1,2)$. Then, the CM can be obtained by $V_{12} = \boldsymbol{R} G \boldsymbol{R}^T$,
where $\boldsymbol{R}$ is the matrix representing the inverse of the coordinate transformations \eqref{eq:r+-}.  It is straightforward to calculate
\begin{equation}
\label{eq:CMSPDC}
V_{12} = \begin{pmatrix}
{A}  & C  \\
 {C}^{T} &  {B} \\
\end{pmatrix},
\end{equation}
where $4 \times 4$ matrices of type $A$ ($B$) refer to photon 1 (2), and the $C$ matrices refer to correlations between the photons.   We have $A=B$, with
\begin{equation}
A = \begin{pmatrix}
\sigma^2+\sigma_{-}^2 & -\frac{k_p \sigma^2}{2 R} & 0 & k u\sigma^2\\
-\frac{k \sigma^2}{2 R} &\frac{1}{4}( \tau^2 + \Delta_-^2) & -k u\sigma^2 & 0 \\
0 & - k u\sigma^2 & \sigma^2+\sigma_{-}^2 & -\frac{k \sigma^2}{2 R} \\
k u\sigma^2 & 0 & -\frac{k \sigma^2}{2 R} &\frac{1}{4}( \tau^2 + \Delta_-^2)
\end{pmatrix}. \label{ABjj}
\end{equation}Thus, there are cross-correlations between the near-field (position) and far-field (momentum) coordinates of each down-converted photon that are proportional to the twist phase $u$. This results in orbital angular momentum that is one-half that of the pump beam: $\langle L_z \rangle  =   \hbar k u\sigma^2$, implying that it is conserved from the pump to the down-converted photons.
\par
The matrices $C$ describe correlations between photons 1 and 2. We have
\begin{equation}
C =  \begin{pmatrix}
\sigma^2-\sigma_{-}^2 & -\frac{k \sigma^2}{2 R} & 0 & k u\sigma^2\\
-\frac{k\sigma^2}{2 R} &  \frac{1}{4}(\tau^2 - \Delta_-^2) & -k u\sigma^2 & 0  \\
0 & -k u\sigma^2 & \sigma^2-\sigma_{-}^2 & -\frac{k \sigma^2}{2 R}  \\
k u\sigma^2 & 0  &-\frac{k\sigma^2}{2 R} &  \frac{1}{4}(\tau^2 - \Delta_-^2)  \\
\end{pmatrix}. \label{Cjj}
\end{equation}
 The wave vector correlations diverge more rapidly due to the presence of the twist phase in the $\tau^2$ term.
We see that $\langle x_1 q_{y2} \rangle - \langle x_2 q_{y1} \rangle =  k u\sigma^2$.  This is not an optical angular momentum per se, but rather a coupling between perpendicular components of the position of one photon and the wave vector of the other \cite{gomes09a}.  
\par
{\em Twist phase and entanglement.---} In typical SPDC experiments, entanglement can be identified by observing correlations in the near-field (position) variables and the far-field (wave vector) variables, leading to violation of one of two inequalities  \cite{mancini02}:

 \begin{equation}
\langle\Delta r_{\pm}\rangle \langle \Delta q_{r \mp}\rangle  \geq \frac{1}{2}
\label{eq:lambda}
 \end{equation}
   where $r_{\pm}=x_{\pm}, y_{\pm}$ and $q_{r \pm}=q_{x \pm}, q_{y \pm}$. Violation of inequalities \eqref{eq:lambda} occurs when either the two-photon state is anti-correlated in the near-field and correlated in the far-field ("+-"), or correlated in the near-field and anti-correlated in the far-field ("-+").  The latter is typically the case in SPDC, and a number of experiments have used these or similar inequalities to identify the entanglement of the down-converted photons \cite{howell04,edgar12,schneeloch19,Defienne19,Zhang19,tasca08}.
    \par
While the above criteria are sufficient for many experiments, they fail to capture entanglement that arises from correlations between different spatial DoF.  A more complete analysis is achieved by calculating the four symplectic eigenvalues $\{ \lambda_i \}$  of the CM of the partially transposed state \cite{adesso07}.   This allows us to investigate so-called distillable Gaussian entanglement. Partial transposition corresponds to changing the sign of the wave vector coordinates of one of the photons \cite{simon00}.   To simplify the analysis, we perform a local scaling of the coordinates of each photon using the transformation $\xi^\prime_{12}=\boldsymbol{S} \xi_{12}$, with $\boldsymbol{S}=\oplus_{k=1}^{4} S_k$ and $S_k=\mathrm{diag}(1/\sqrt{2},\sqrt{2})$. In this case, observing $\lambda_i < 1/2$ implies that the state has a negative partial transpose, which indicates  entanglement \cite{adesso07,simon00}.  Furthermore, the smallest symplectic eigenvalue can be used to quantify gaussian entanglement using the negativity or other quantifiers \cite{adesso07}.  The symplectic eigenvalues of \eqref{eq:CMSPDC} are two-fold degenerate, and given by (see Appendix \ref{supmat})
 \begin{equation}
\lambda_\pm = \frac{1}{\sqrt{2}} \left| \sqrt{a_+ \pm \sqrt{4 k^2 \Delta_-^2 \sigma_-^2 \sigma^4 \left(u^2+\frac{1}{R^2}\right)+a_-^2}}\right|,
 \end{equation}
 where $a_\pm= \tau^2 \sigma_-^2 \pm \Delta_-^2 \sigma^2$. 
\begin{figure}
\centering
\includegraphics[width=6cm]{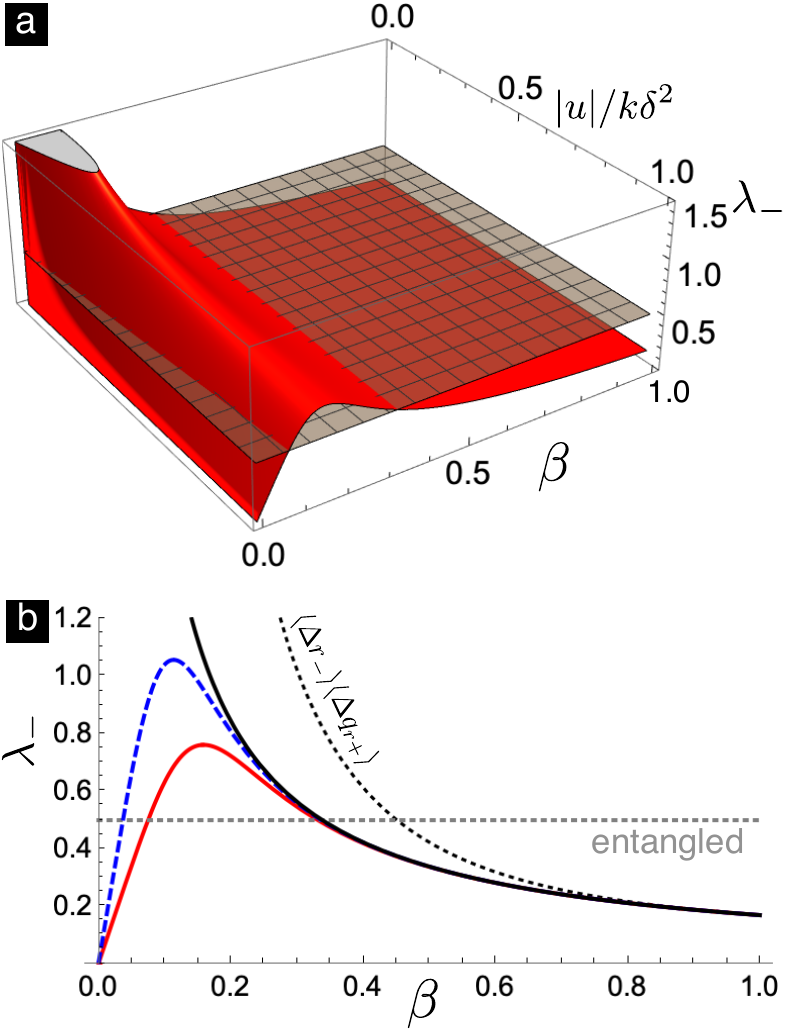}
\caption{a) Evaluation of smallest symplectic eigenvalue $\lambda_-$ (red surface) as a function of the normalized pump beam coherence $\beta$ and normalized twist phase.  Entanglement is confirmed when $\lambda_- < 1/2$ (gray horizontal plane). The SPDC parameters are $R=\infty$, $\lambda_p=400$nm, $\sigma_p=50 \mu$m, and $L=1$cm. b) Profile plots of $\lambda_-$ for normalized twist phase $|u|/k \delta^2$ equal to zero (black solid line), 1 (red solid line) and 1/2 (blue dashed line).  The dotted black curve is the near-field/far-field entanglement criteria \eqref{eq:lambda}.}
\label{fig:3dplots}
\end{figure}

We can now analyze the entanglement produced in the SPDC process as a function of the spatial coherence of the TGSM pump beam and its twist phase.  Fig. \ref{fig:3dplots} a) shows a plot of the smallest symplectic eigenvalue (red surface) as a function of the absolute value of the twist phase $|u|$ and the spatial coherence parameter $\beta$ of the TGSM pump beam.  Here, $|u|$ is scaled by $1/k \delta^2$, so that it varies from 0 to 1.  Two regions of entanglement are clearly visible.  One corresponds to larger coherence parameter $\beta$.  This is the usual case, where entanglement increases as a function of the pump beam coherence \cite{Giese18,Defienne19,Zhang19}.  However, we also observe a second region where entanglement grows inversely with the coherence, and is present even though $\beta$ nears zero.  We note that this only occurs for larger values of the twist phase.

To better visualize these results, in Fig. \ref{fig:3dplots} b) we plot $\lambda_-$ as a function of $\beta$ for  $|u|/k \delta^2=0,1/2,1$. In all cases we can identify entanglement when the pump beam is more coherent.    Also shown is the LHS of of inequality \eqref{eq:lambda}, showing violation for larger $\beta$.  This indicates that for larger pump coherence, the entanglement involves the same spatial DoF of the photons, and can be detected using the standard approach. Indeed, for a coherent pump beam, with $R=\delta=\infty$ and twist phase $u=0$, the symplectic eigenvalues $\lambda_{\pm}$ reduce to the LHS of entanglement criteria \eqref{eq:lambda}.  Nevertheless, Fig.  \ref{fig:3dplots} b) shows that this "standard" entanglement decreases and eventually disappears as the pump beam becomes more incoherent, as observed in Refs. \cite{Giese18,Defienne19,Zhang19}. 
 \par
For non-zero twist phase, our study reveals that a different type of entangled quantum state can be generated for small values of $\beta$. Here the two-photon state is highly mixed, since the purity $\mu_{12} \propto \beta^2$.   Interestingly, the entanglement can be larger when the pump beam is less coherent.
In this region, there is no violation of the near-field/far-field criteria \eqref{eq:lambda}.  To identify entanglement one can measure the elements of the CM using optical transformations \cite{tasca08}, or use the optimization techniques presented in Ref. \cite{hyllus06} to find more efficient entanglement criteria. 
\par
To provide an intuitive explanation for this phenomenon, note that when $u \neq 0$, the correlation matrix $C$ contains extra covariance terms with modulus $\propto k \sigma^2 |u|$, which are upperbounded by $\sigma^2/\delta^2$.  These covariances originate from the correlations already present in the pump beam, which transformed into quantum correlations between the down-converted photons. When the twist phase is appreciable, the correlations grow at the same rate that the purity decreases ($\mu_{12} \propto \beta^2 \sim \delta^2/\sigma^2$).   On the other hand, when $u=0$, these covariances are zero, demonstrating that this is an effect that could only be revealed while considering this general class of TGSM beams in the SPDC process.   The existence of highly mixed yet highly entangled states is a known phenomenon that appears for infinite dimensional systems (for another example see Ref. \cite{adesso05}), since for these systems there is no dense region of separable states in the state space \cite{clifton99,eisert02}. Thus, there exist entangled mixed states that lie arbitrarily close to separable mixed states.  We note that there is usually some physical (e.g. energy) constraint in the generation of infinite dimensional systems.   In our case, the relevant physical parameter is the number of transverse modes supported by the optical systems, providing a lower limit to $\beta$.  We note that SPDC experiments have been realized with $\beta \lesssim 0.1$ \cite{Zhang19}, indicating that experimental observation of this phenomenon is feasible with current technology.

{\em Conclusions.---} In this work we bridge together the more general theory of partially coherent Gaussian beams, namely the twisted Gaussian-Schell model, and the process of spontaneous parametric down-conversion that has been used extensively over the last decades to produce entangled photons. By doing so, we reveal new phenomena that allow for the generation of a class of multi-photon states with unique entanglement and coherence properties. Even though similar entangled states might be created in quantum optics of entangled qumodes \cite{adesso05,vanenk03}, here we are able to interconnect the amount of entanglement with the the so-called twist phase of the pump beam, an intriguing optical phenomenon first introduced in 1993 \cite{simon93}.  We note that twist phase is a property of incoherent Gaussian beams that vanishes in the coherent limit. Thus, the novel entanglement produced here is directly related to the  incoherence of the pump beam.    We can envisage a number of potential applications, and we expect that these highly mixed yet highly entangled states should allow for the exploitation of highly entangled photons in quantum adaptations of applications originally designed for incoherent beams, such as imaging and optical communications.   For example, a very recent study has shown that partially coherent multi-photon states are more resistant to atmospheric turbulence \cite{phehlukwayo2020}.   Our results should provide a way to increase the spatial entanglement in this scenario.    
\begin{acknowledgements}
SPW is grateful to R. Simon for valuable discussions during his visit to UFRJ.  The authors would like to thank P.H. Souto Ribeiro, D.S. Tasca and E. G\'omez for helpful comments, and the Chilean agencies Fondo Nacional de Desarrollo Cient\'{i}fico y Tecnol\'{o}gico (FONDECYT) (1200266, 1200859); Millennium Institute for Research in Optics (MIRO), and the Brazilian agencies CAPES, CNPQ and the INCT-IQ for partial financial support.  This work was realized as part of the CAPES/PROCAD program.
\end{acknowledgements}

\appendix
\section{Partially coherent SPDC}
\label{supmat}
Here we construct the two-photon state using a coherent mode decomposition of the partially coherent pump beam.  Alternative treatments of SPDC with a partially coherent pump beam can be found in Refs. \cite{Giese18,Defienne19,Zhang19}. 
\subsection{Brief Review of coherent SPDC}
For a coherent pump laser beam, it is well known that the state describing the spatial degrees of freedom of photon pairs produced from SPDC in a thin non-linear crystal can be well approximated by  \cite{hong85,monken98a,walborn10,Schneeloch16}
\begin{equation}
\ket{\psi}= \iint  d\boldsymbol{q}_1 d\boldsymbol{q}_2 \phi (\boldsymbol{q}_1,\boldsymbol{q}_2) \ket{\boldsymbol{q}_1} \ket{\boldsymbol{q}_2}, 
\label{eq:cspdc}
\end{equation}
where $\boldsymbol{q}_l$ ($l=1,2$) are transverse components of the down-converted wave vectors.  The state  $\ket{\boldsymbol{q}_l}$ represents a single photon with transverse wave vector $\boldsymbol{q}_l$ and frequency $\omega_l$.  Here it is assumed that the photons are nearly monochromatic, so that $\Delta \omega \ll \omega$, which can be guaranteed by using narrowband frequency filters.  When the  down-converted photons are degenerate, $w_1 \approx w_2 \approx w/2$,  the two-photon amplitude can be written as a product of functions of the sum and difference coordinates of the down-converted photons:
\begin{equation}
\phi (\boldsymbol{q}_1,\boldsymbol{q}_2)= v(\boldsymbol{q}_1+\boldsymbol{q}_2) \mathcal{S} \left(\left[ \boldsymbol{q}_1-  \boldsymbol{q}_2 \right]^2 \right ),
\label{eq:phipump2}
\end{equation}
 where $\mathcal{S}(x)=\mathrm{sinc}(L x/ 4 k)$, and $L$ is the length of the non-linear crystal.   Here the paraxial approximation $k_z  \approx  k (1-{q^2}/{2k^2})$ is assumed for all fields.   The crystal is assumed to be thin enough so that walk-off effects from the birefringence can be ignored. 
 
 \subsection{Density operator for partially coherent SPDC}
We can write the density operator for partially coherent SPDC as a convex sum of pure states $\ket{\psi_j}$, identifying each $\ket{\psi_j}$ with a different coherent SPDC process of the form \eqref{eq:cspdc}, labelled by $j$, and described by the two-photon amplitude function $\phi_j$: 
\begin{equation}
\varrho=\sum_j c_j	\ket{\psi_j}\bra{\psi_j}, 
\label{eq:rho}
\end{equation}
where the real coefficients satisfy $c_j \geq 0$ and $\sum_j c_j = 1$.   The density matrix \eqref{eq:rho} is completely general.  Below, we we consider that the pump beam is spatially coherent, and that it pumps a single down-conversion crystal.  Then, the coefficients $c_j$ are determined by the incoherent sum representation of the incoherent pump beam.   We note that the incoherent sum representation of coherent beams has been used to produce twisted Gaussian Schell beams, as in \cite{friberg94}. 
\subsection{Wigner function}
 Using the general form of the density operator  \eqref{eq:rho} and the two-photon state for coherent SPDC \eqref{eq:cspdc},  the Wigner function for partially coherent SPDC is 
\begin{equation}
\mathcal{W}(\boldsymbol{r}_1,\boldsymbol{q}_1,\boldsymbol{r}_2,\boldsymbol{q}_2) = \sum_j c_j W_j(\boldsymbol{r}_1,\boldsymbol{q}_1,\boldsymbol{r}_2,\boldsymbol{q}_2), 
\label{eq:W}
\end{equation}
where $W_j$ is the Wigner function for each of the pure two-photon states in \eqref{eq:rho}. To simplify this expression, we can define the global variables
\begin{align}
\boldsymbol{q}_{\pm} &=\boldsymbol{q}_1\pm\boldsymbol{q}_2 \label{eq:q+-} \\
{\boldsymbol{r}}_{\pm} &= \frac{1}{2} (\boldsymbol{r}_1 \pm \boldsymbol{r}_2) \label{eq:r+-} 
\end{align}
and rewrite the Wigner function \eqref{eq:W} in terms of the global coordinates as 
\begin{equation}
\mathcal{W}(\boldsymbol{r}_+,\boldsymbol{q}_+,\boldsymbol{r}_-,\boldsymbol{q}_-) = \sum_j c_j W_{j+}(\boldsymbol{r}_+,\boldsymbol{q}_+)  W_{j-}(\boldsymbol{r}_-,\boldsymbol{q}_-), 
\label{eq:Wiggen}
\end{equation}
where
\begin{align}
W_{j+}(\boldsymbol{r}_+,\boldsymbol{q}_+)  =& \frac{1}{(2\pi)^2} \iint d\boldsymbol{Q}_+ 
\operatorname{e}^{i {\boldsymbol{r}}_+\cdot\boldsymbol{Q}_+} \nonumber \\
& v_j \left(\boldsymbol{q}_+ + \frac{1}{2} \boldsymbol{Q}_+ \right ) \nonumber \\ 
& {v_j}^* \left(\boldsymbol{q}_+ - \frac{1}{2} \boldsymbol{Q}_+\right ), 
\end{align}
and
\begin{align}
W_{j-}(\boldsymbol{r}_-,\boldsymbol{q}_-)  =&\frac{1}{(2\pi)^2} \iint  d\boldsymbol{Q}_- \operatorname{e}^{i{\boldsymbol{r}}_-\cdot\boldsymbol{Q}_-} \times \nonumber \\
&\mathcal{S}_j\left(\left [\boldsymbol{q}_- +  \frac{1}{2}\boldsymbol{Q}_-\right]^2\right) \times  \nonumber \\
& \mathcal{S}_j^*\left( \left[\boldsymbol{q}_- -  \frac{1}{2} \boldsymbol{Q}_-\right]^2\right). \label{Wigner+-Paraxial}
\end{align}
As mentioned above, we will now suppose that a single down-conversion crystal is pumped by a beam with non-unity degree of spatial coherence. Then we have $W_{j-}(\boldsymbol{r}_-,\boldsymbol{q}_-) = W_{-}(\boldsymbol{r}_-,\boldsymbol{q}_-)$ for all $j$, and we can write
 \begin{equation}
\mathcal{W}(\boldsymbol{r}_+,\boldsymbol{q}_+,\boldsymbol{r}_-,\boldsymbol{q}_-) =W_{-}(\boldsymbol{r}_-,\boldsymbol{q}_-)  \sum_j c_j W_{j+}(\boldsymbol{r}_+,\boldsymbol{q}_+), 
\label{eq:W2}
\end{equation}
that is the total Wigner function factorizes into functions involving the two sets of global variables.  In this scenario, we can recognize the function corresponding to the $(\boldsymbol{r}_+,\boldsymbol{q}_+)$ variables as an incoherent sum of Wigner functions $W_{j+}$ corresponding to the profiles of coherent laser beams \cite{bastiaans86}. That is, the coefficients $c_j$ and  functions $W_{j+}$ are determined by the incoherent sum representation of the pump beam.  The two-photon Wigner function can be written as 
 \begin{equation}
\mathcal{W}(\boldsymbol{r}_+,\boldsymbol{q}_+,\boldsymbol{r}_-,\boldsymbol{q}_-) =W_{-}(\boldsymbol{r}_-,\boldsymbol{q}_-)   W_{+}(\boldsymbol{r}_+,\boldsymbol{q}_+), 
\label{eq:WpumpPC}
\end{equation}
where $W_{+}(\boldsymbol{r}_+,\boldsymbol{q}_+)$ represents the summation term in \eqref{eq:W2}. 
\subsection{Purity of the two-photon state}
The purity of a quantum state can be determined from the Wigner function \cite{adesso07}.  For two photon states, the purity is given by
\begin{equation}
\mu_{12} = \iiiint d\boldsymbol{r}_+ d\boldsymbol{q}_+ d\boldsymbol{r}_- d\boldsymbol{q}_- \mathcal{W}(\boldsymbol{r}_+,\boldsymbol{q}_+,\boldsymbol{r}_-,\boldsymbol{q}_-)^2.
\end{equation}
When the Wigner function factorizes as in Eqs. \eqref{eq:WpumpPC}, the purity also factorizes: $\mu_{12}=\mu_+ \mu_-$, where
\begin{equation}
\mu_{\pm} = \iint d\boldsymbol{r}_\pm d\boldsymbol{q}_\pm {W}_\pm(\boldsymbol{r}_\pm,\boldsymbol{q}_\pm)^2. 
\end{equation}
Since the Wigner function $W_+$ is that of the pump beam, the purity $\mu_+$ can be shown to be equivalent to the transverse spatial coherence of the beam.  Thus, the two-photon purity is directly proportional to the coherence of the pump laser. 
\section{Partial Transposition and Symplectic Eigenvalues}
The partial transposition of a bipartite state is obtained by inverting the sign of the momentum coordinate of one of the systems \cite{simon00}--say system 2.   In our case, this corresponds to the transformation $\boldsymbol{\xi}_2 \rightarrow \boldsymbol{T}\boldsymbol{\xi}_2$, where $\boldsymbol{T}=(1,-1,1,-1)$.  The covariance matrix of the partially transposed state can be calculated by $V^{PT}=(\boldsymbol{I} \oplus \boldsymbol{T}) V(\boldsymbol{I} \oplus \boldsymbol{T})$.  The symplectic eigenvalues of $V^{PT}$ can be obtained by calculating the eigenvalues of the matrix $|i \Omega V^{PT}|$, where the symplectic form ${\Omega}=\bigoplus_{j=1}^{4} \omega$ and 
\begin{equation}
\omega = 
 \begin{pmatrix}
0 & 1 \\
-1 & 0
\end{pmatrix}.
\end{equation}      
Details on gaussian entanglement and quantum information with gaussian states can be found in Refs. \cite{braunstein05,adesso07}.



\end{document}